\documentclass[aps,prl,twocolumn,groupedaddress,showpacs]{revtex4}
\usepackage{graphicx,amsmath}

\newcommand{\Pe}{\mbox{P\hspace{-1pt}e\hspace{1pt}}}
\newcommand{\vz}{\mathrm{v}\hspace{-1pt}_z}
\newcommand{\vrel}{\mathrm{v}\hspace{-1pt}_r}

\newcommand{\dt}{\Delta t}
\newcommand{\dx}{\Delta x}
\newcommand{\dz}{\Delta z}

\newcommand{\dxsq}{\langle\Delta x^2\rangle}
\newcommand{\dxqd}{\langle\Delta x^4\rangle}
\newcommand{\dzsq}{\langle\Delta z^2\rangle}
\newcommand{\sx}{\sigma\hspace{-1pt}_x}
\newcommand{\sz}{\sigma\hspace{-1pt}_z}

\begin{document}

\title{Diffusion and mixing in gravity-driven dense granular flows}
\author{Jaehyuk Choi$^1$, Arshad Kudrolli$^2$, Rodolfo R. Rosales$^1$ and Martin Z. Bazant$^1$}
\affiliation{
  $^1$ Department of Mathematics, Massachusetts Institute of Technology, Cambridge, MA 01239 \\
  $^2$ Department of Physics, Clark University, Worcester, MA 01610}
\date{\today}

\begin{abstract}
We study the transport properties of particles draining from a silo
using imaging and direct particle tracking. The particle
displacements show a universal transition from super-diffusion to
normal diffusion, as a function of the distance fallen, independent of
the flow speed. In the super-diffusive (but sub-ballistic) regime,
which occurs before a particle falls through its diameter, 
the displacements
have fat-tailed and anisotropic distributions. In the diffusive
regime, we observe very slow cage breaking and P\'eclet numbers of
order 100, contrary to the only previous microscopic model (based on
diffusing voids). Overall, our experiments show that diffusion and
mixing are dominated by geometry, consistent with fluctuating contact
networks but not thermal collisions, as in normal fluids.
\end{abstract}


\maketitle

Granular flow is an attractively simple and yet surprisingly complex
subject~\cite{jaeger96}. Fast, dilute flows are known to obey
classical hydrodynamics (with inelastic collisions), but slow, dense
flows pose a considerable challenge to theorists, due to many-body
interactions and non-thermal fluctuations.  Beyond their fundamental
scientific interest, such flows have important engineering
applications~\cite{ottino02}, e.g. to new pebble-bed nuclear
reactors~\cite{talbot02}, whose efficiency and safety depend on the
degree of mixing in very slow granular drainage ($<1$ pebble/min).

Although dense granular drainage is very familiar (e.g. sand in an
hourglass), it is far from fully understood. 
Over the past forty years, a number of theoretical approaches have been 
proposed for steady state flow~\cite{lit63,mullins72,nedderman79,prakash91,nedderman91}.
Continuum approaches are based on the critical-state theory of soil mechanics
and yield only mean velocity fields~\cite{prakash91,nedderman91}.
On the other hand, the diffusive void model~\cite{lit63,mullins72} 
takes a particle approach, in
which `voids' injected at the orifice cause drainage by diffusing
upward and exchanging position with particles along the way. Averaging over the
void trajectories yields the same continuum velocity field for
particles as the `kinematic model'~\cite{nedderman79,nedderman91},
which provides a reasonable fit to experimental data with only one
fitting parameter (the diffusion length, $b$)
~\cite{mullins72,mullins74,nedderman79,samadani99}, 
although the void model on a regular lattice (as in
Ref.~\cite{caram91}) underpredicts its
value~\cite{bazant03}. Remarkably, these models depend only on {\it
geometry} and not on momentum, energy, etc.

In spite of the success of the kinematic model, however, its only
microscopic basis, the void model, greatly overpredicts diffusion. To
see this, consider the P\'eclet number, $\Pe_x = \vz d / D_x$, the
dimensionless ratio of advection in uniform downward flow of speed,
$\vz$, to diffusion with a horizontal diffusivity, $D_x$, at the scale
of a particle diameter, $d$. In the void model, when a particle falls a
distance $\Delta h$, $\Pe_x
= (\Delta h/\dt) d / (\dxsq/2 \dt) = \Delta h\, 2d / \dxsq$, is of
order one for any conceivable packing since $\Delta h \approx \dx
\approx d$, and therefore it diffuses horizontally by roughly 
$\sqrt{\Delta h}$. This prediction is
contradicted by everyday experience and our experiments below, which
exhibit far less mixing.  An attempt to resolve this paradox with a
new model appears in a companion paper~\cite{bazant03}.

\begin{figure}
  \parbox{0.43\linewidth}{
  \hbox{\hbox to 0pt{(a)\hss}\vtop{\hbox{}\vspace{-1em}
    \includegraphics[width=0.75\linewidth]{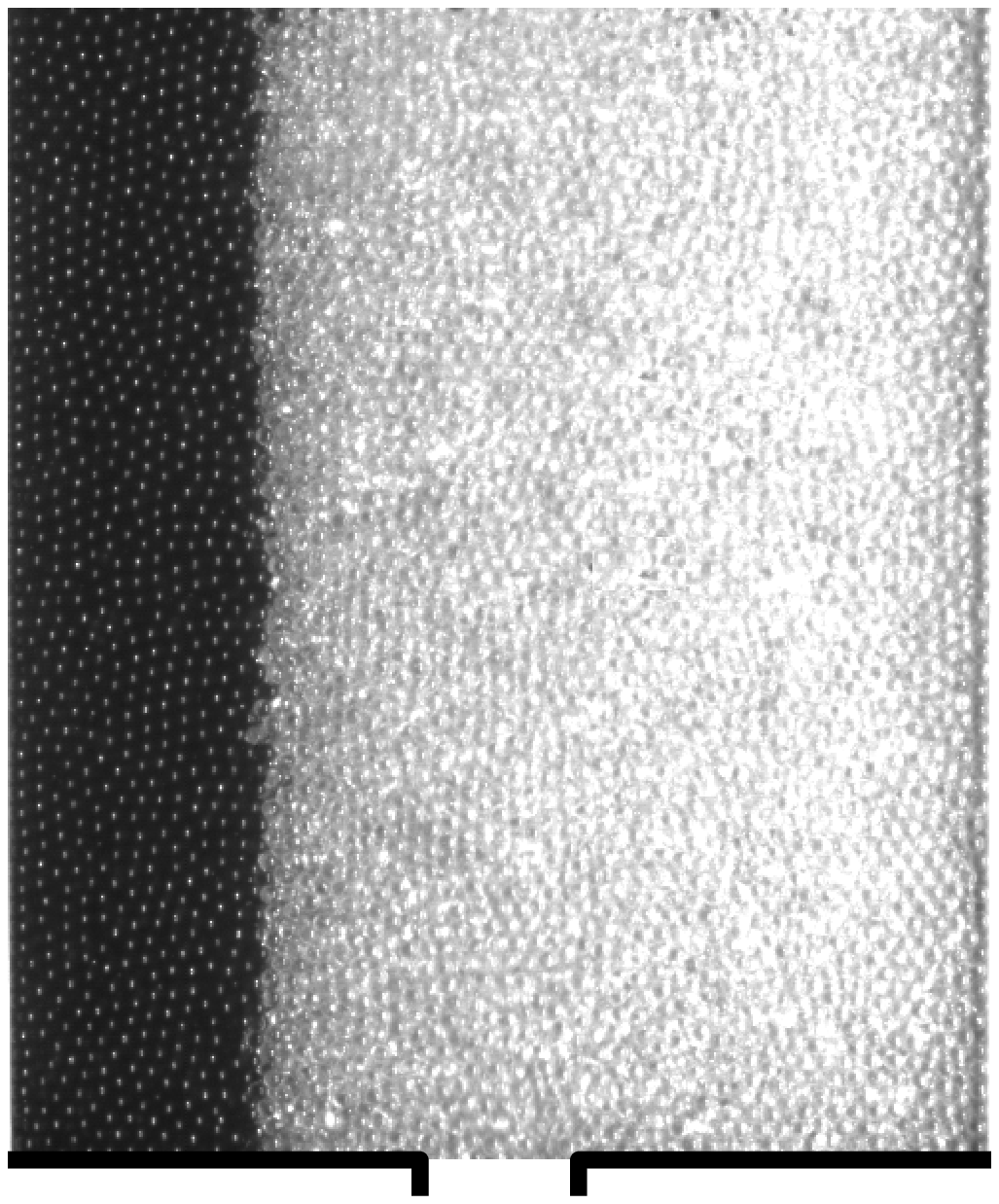}}}\vspace{2mm}
  \hbox{\hbox to 0pt{(b)\hss}\vtop{\hbox{}\vspace{-1em}
    \includegraphics[width=0.75\linewidth]{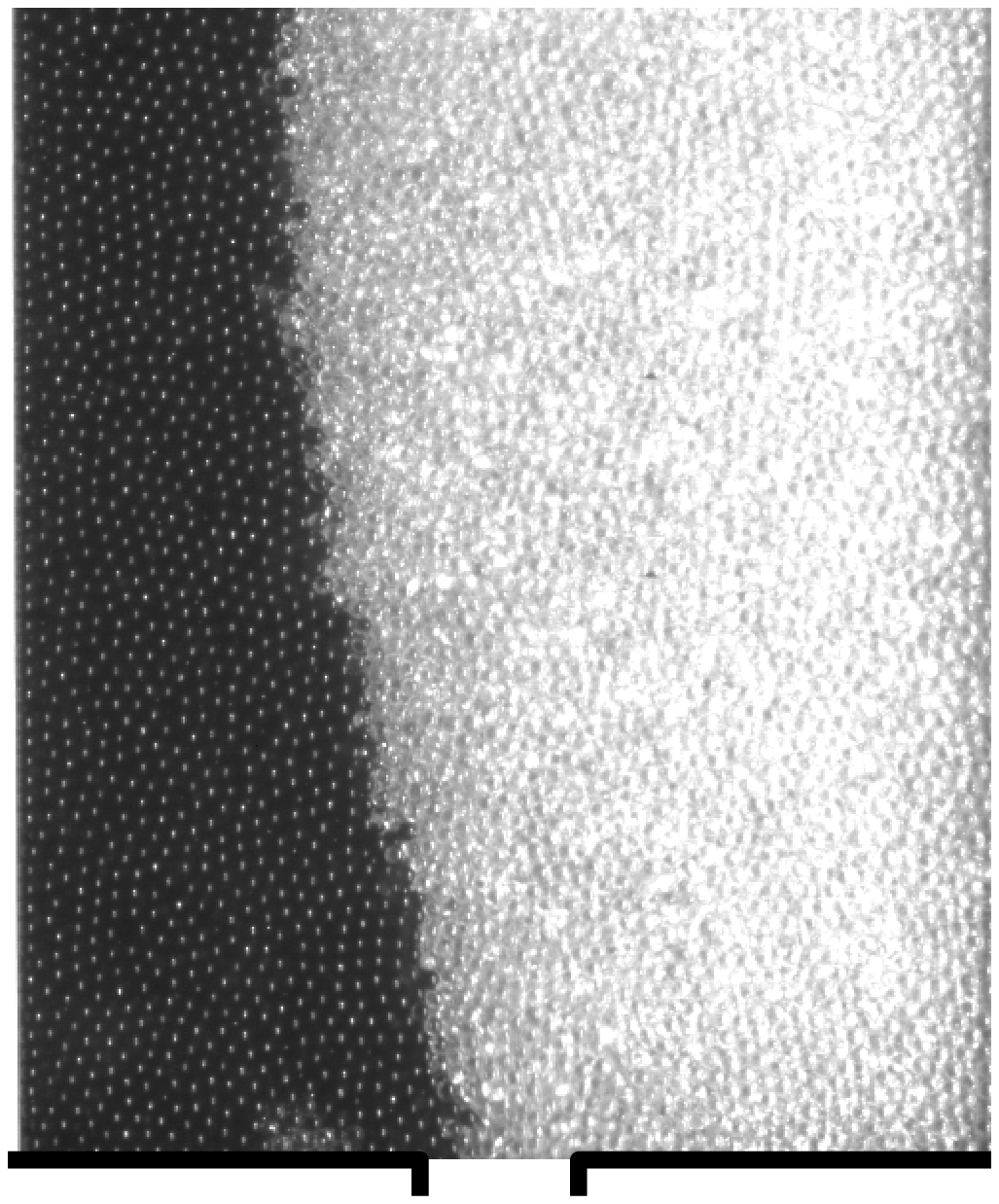}}} }
  \parbox{0.54\linewidth}{
  \hbox{\hbox to 0pt{(c)\hss}\vtop{\hbox{}\vspace{-1em}
    \includegraphics[width=0.85\linewidth]{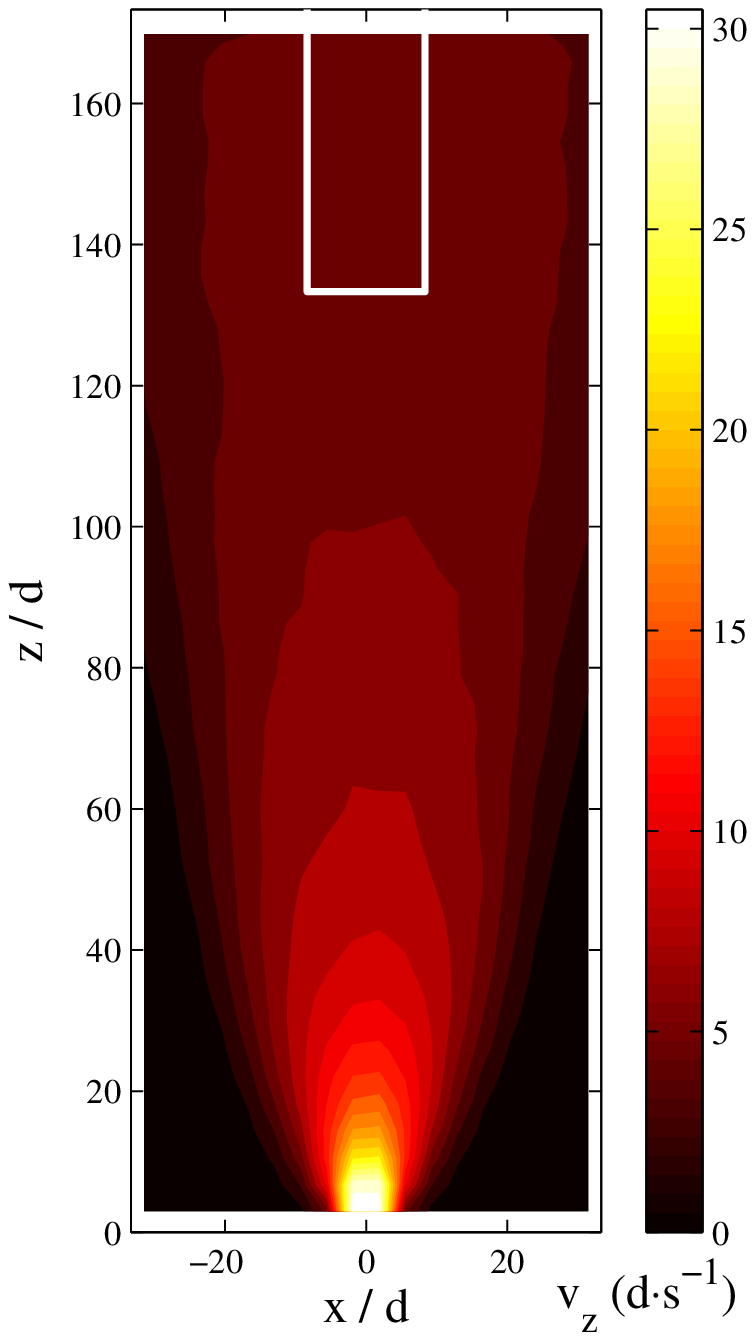}}} }
  \caption{\label{Fig1} An initially flat, off-center interface
    between two regions of differently colored beads (a) stretches and
    roughens after draining half of the silo (b), but little mixing is
    observed. (c) Contour plot of the average downward velocity field,
    $\vz\/$, with an orifice width, $W=16$ mm = $5.3$~$d$, where $d$
    is the particle diameter.  The white box indicates a region of
    nearly uniform flow where all subsequent measurements are made.}
\end{figure}

In this Letter, we describe particle-tracking experiments on silo
drainage using similar techniques as in a recent (lower-resolution)
study of the velocity field~\cite{medina98}. We focus on the
statistical evolution of particle displacements and topological
``cages'', which should aid in developing new microscopic models.  Our
data may also have implications for recent attempts to apply
thermodynamic approaches from glassy dynamics to granular
flows~\cite{edwards94,barrat00,makse02}. Although we do not define a
``granular temperature'', we observe the presumably related effect of
varying the flow rate in all of our measurements.

Our experimental apparatus involves glass beads ($d=3.0 \pm 0.1$ mm)
in a quasi-two dimensional silo ($20.0 \times 90.0 \times 2.5$ cm $=
67 \times 300 \times 8.3$ $d$). The particles are observed near the
front wall made of transparent glass, where the slight polydispersity
reduces the tendency for hexagonal packing. (As seen in
Fig.~\ref{Fig1}, there is no long-range order, although the wall induces
some short-range order that may affect our results.) 
We find that varying the thickness of the silo has insignificant effect
on the diffusion properties and therefore we report our data for a single
thickness.
We track individual particles using a high-speed digital camera with 
a maximum resolution of $512\times 1280\/$ pixels at 1000 frames/sec. 
Particle positions are obtained to sub-pixel accuracy using a centroid 
technique ($\pm 0.003 d$, $d = 15$ pixels).
 
Our first experiment provides a visual demonstration that particles
mix much less than predicted by the void model. We load the silo with
black and white (but otherwise identical) glass beads, forming two
separate columns, as shown Fig.~\ref{Fig1}(a). From
Fig.~\ref{Fig1}(b), where half of the particles have drained (in 30
sec), it is clear that the black-white interface has not smeared
significantly, although it has roughened. The small degree of mixing
is consistent with the segregation of bi-disperse beads in a similar
apparatus~\cite{samadani99}.

The mean downward velocity field, $\vz\/$, is shown in
Fig.~\ref{Fig1}(c) throughout the silo, as measured by direct particle
tracking.  A distributed filling procedure similar to that in 
Ref.~\cite{zhong01} was used to add grains to the silo. Then the orifice
was opened and steady state flow was allowed to develop before acquiring 
the data used for subsequent analysis.
The flow speed is highest at the center-line, with a maximum
near the orifice, and decays to zero toward the sides. Fairly good
agreement with the kinematic model is obtained with $b=1.3 d$.

In order to investigate particle dynamics in a simple setting, we
focus on a small $17 \times 87$ $d$ region of nearly uniform flow, the
white box in Fig.~\ref{Fig1}(c). For the measurements below, we track
about 1370 particles through this window for 4 seconds in steps of 1
ms. The average flow speed $\vz\/$, the only control parameter in this
study, is varied by changing the width of the orifice, $W\/$. The flow
is fairly smooth for $W \ge 8$ mm (about $3d\/$) and nearly continuous
for $W\ge 16$ mm. For simplicity, we vary $W\/$ in the range 8 mm $\le
W \le\/$ 32 mm, in increments of 4 mm, to avoid the complicated regime
of intermittent flow~\cite{liu98}. This corresponds to
$1.38\;d/\mbox{sec} \le \vz \le 18.39\;d/\mbox{sec}\/$, and data is
consistent with $\vz\propto (W-d)^{1.5}\/$. We compile statistics from
all tracked particles in six experiments per flow speed (except two
for $W=32$mm).

\begin{figure}
  \hbox{\hbox to 0pt{\qquad(a)\hss}\vtop{\hbox{}\vspace{-1em}
      \includegraphics[width=0.7\linewidth]{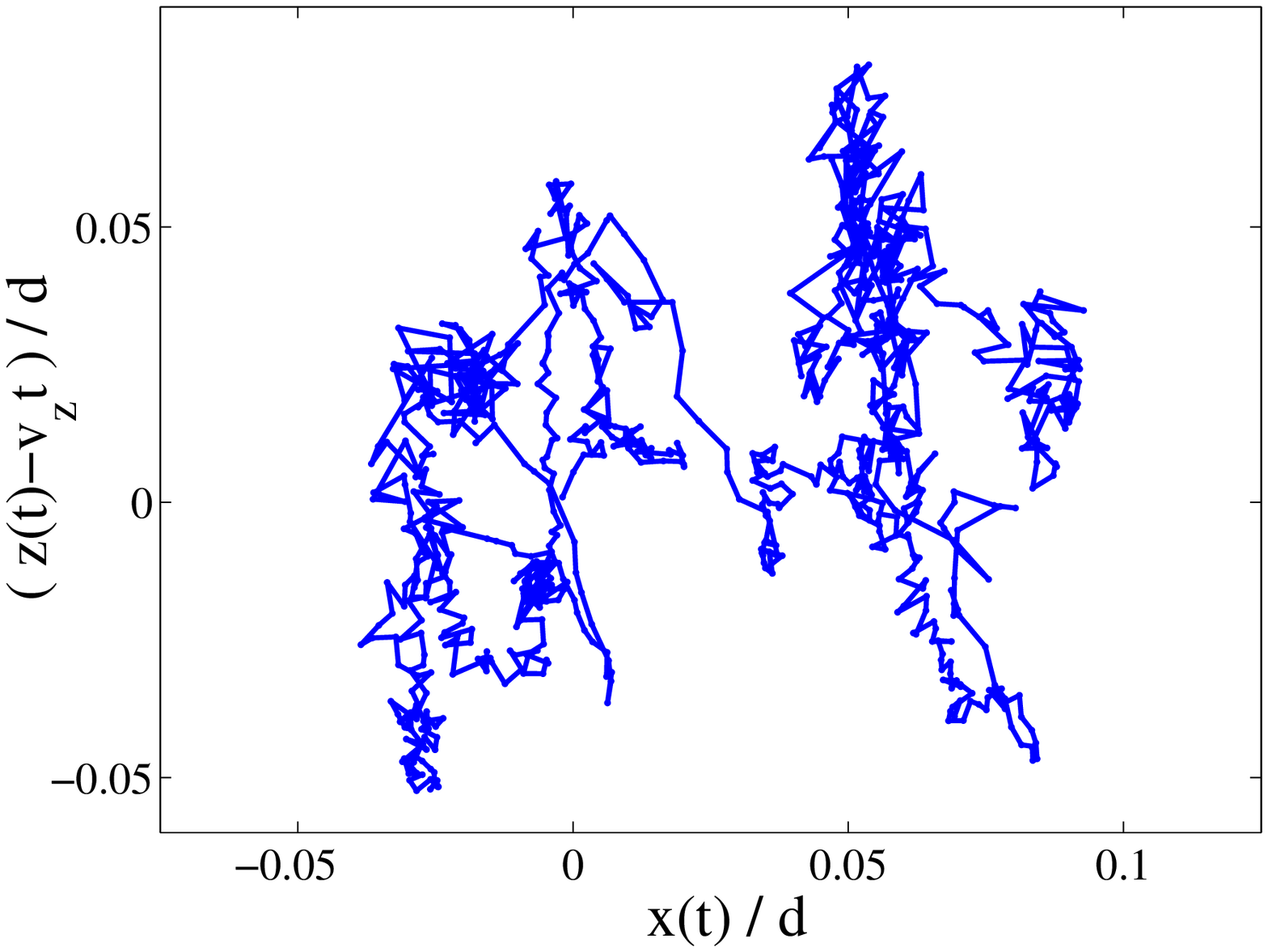}}}
  \hbox{\hbox to 0pt{\qquad(b)\hss}\vtop{\hbox{}\vskip-1em
      \includegraphics[width=0.7\linewidth]{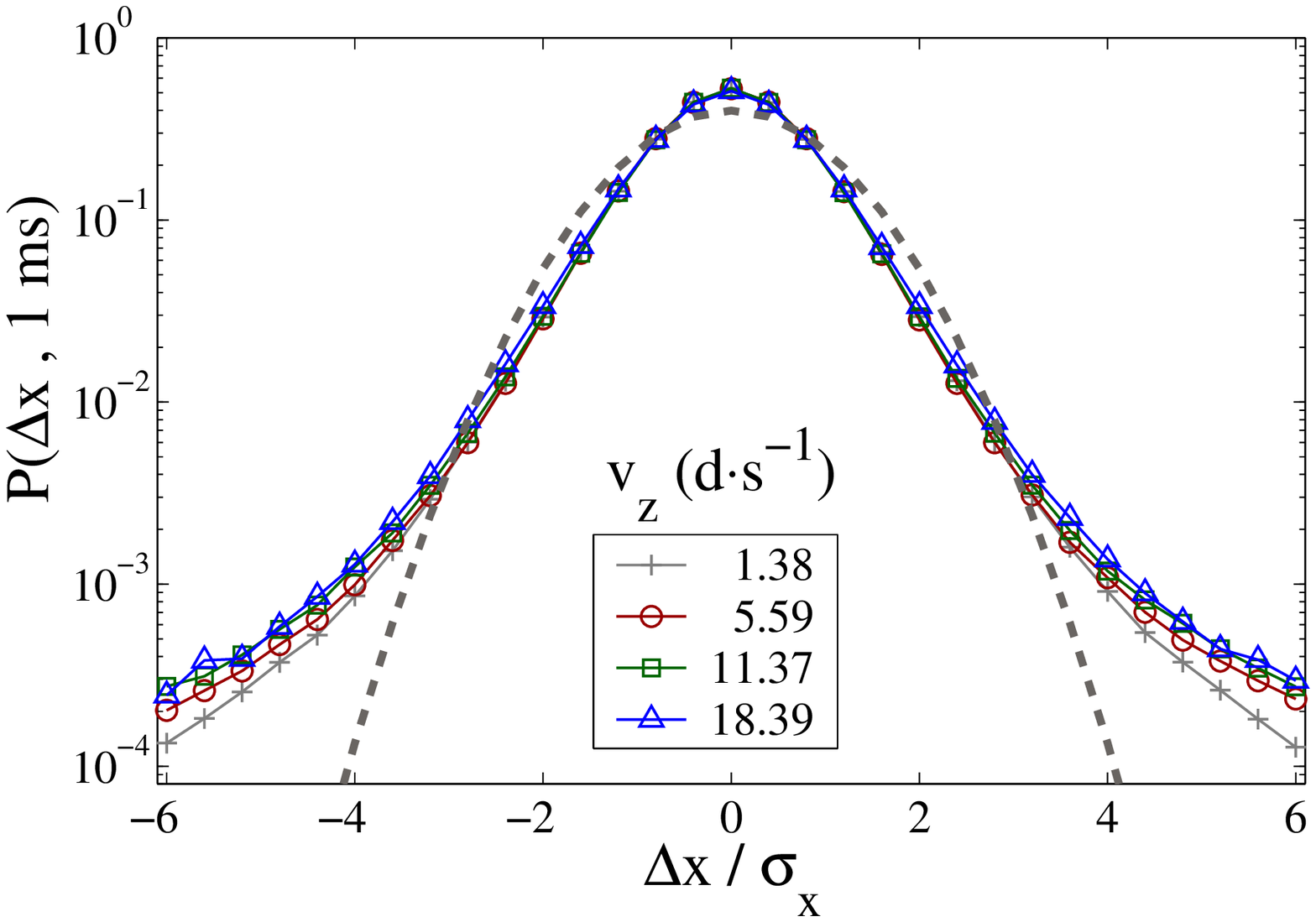}}}
  \hbox{\hbox to 0pt{\qquad(c)\hss}\vtop{\hbox{}\vskip-1em
      \includegraphics[width=0.7\linewidth]{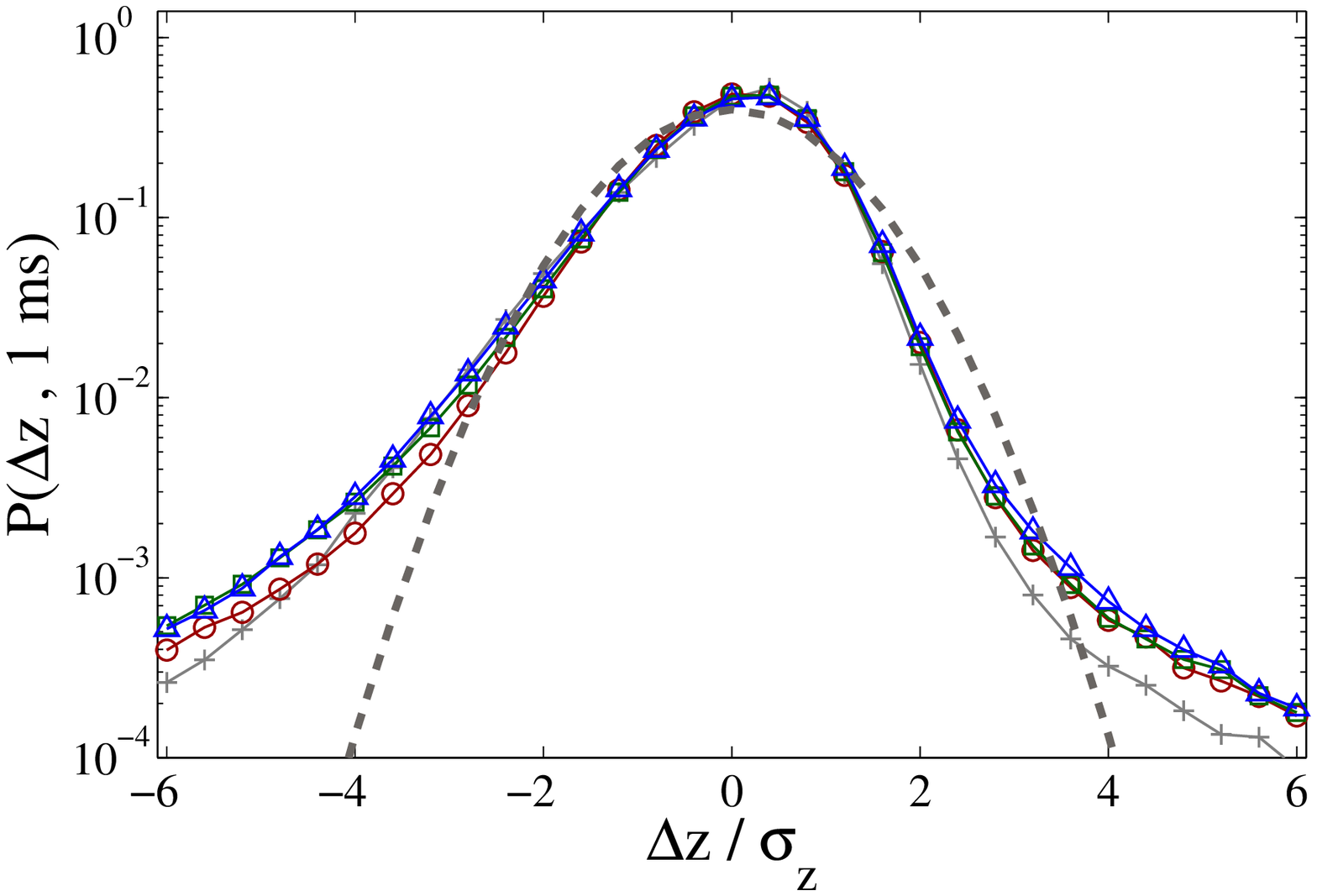}}}
  \hbox{\hbox to 0pt{\qquad(d)\hss}\vtop{\hbox{}\vskip-1em
      \includegraphics[width=0.7\linewidth]{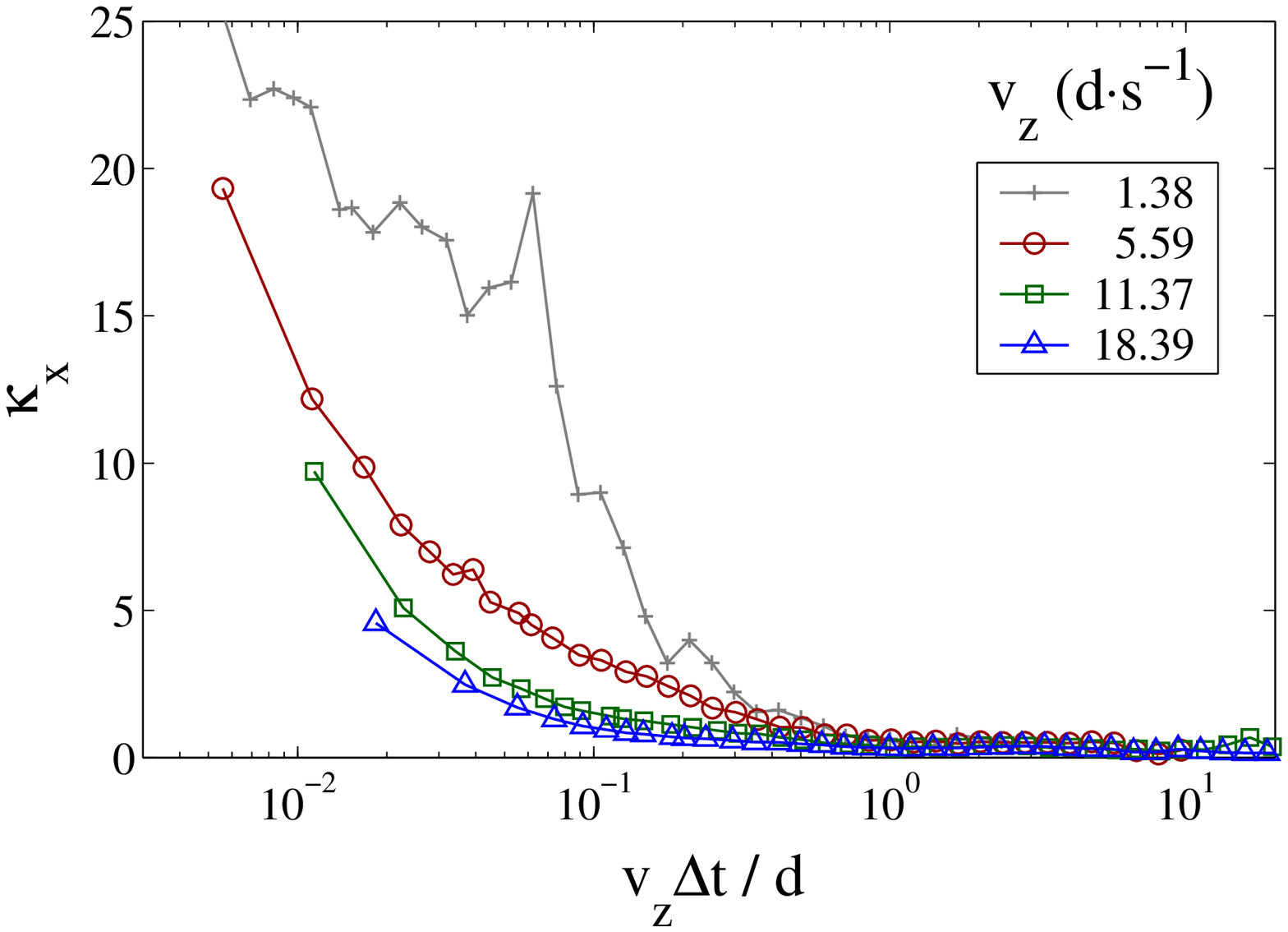}}}
  \caption{\label{Fig2}
    (a) A typical particle trajectory sampled at 1
    ms intervals in a frame moving with the average flow speed,
    $\vz\/$.  (b)--(c) Normalized PDFs for the 1 ms particle
    displacements, $\dx\/$ and $\dz\/$, for various flow speeds,
    $\vz\/$, compared to a standard Gaussian distribution (dotted
    line); standard deviations, $\sx$ and $\sz$, are of order $10^{-3}
    d$. (d) The kurtosis of $\dx$ versus the mean distance fallen,
    $\vz \dt$.}
\end{figure}

From the positions of the particles sampled at 1 ms intervals, we calculate the
horizontal and vertical displacements, $\dx$ and $\dz$, relative to a
frame moving with the mean speed of the flow. A typical trajectory computed in
this way in Fig.~\ref{Fig2}(a) shows periods of small fluctuations with
occasional, much larger steps. This suggests that the probability density
functions (PDFs) for $\dx$ and $\dz$ (for $\dt = 1$ ms) should have fat
tails compared to a Gaussian, which is confirmed in
Fig.~\ref{Fig2}(b)--(c).

Fat-tailed PDFs have also been observed in colloidal glasses and
attributed to cage-breaking~\cite{weeks02}, but a special feature here
is the asymmetry of the PDF for $\dz$ in Fig.~\ref{Fig2}(c). Downward
fluctuations ($\dz < 0$) are larger than both upward ($\dz > 0$) and
horizontal ($\dx$) fluctuations. We attribute this to the fact that
particles are accelerated downward by gravity while being scattered in
other directions by dissipative interactions with neighbors.

\begin{figure}
  \hbox{\hbox to 0pt{\qquad(a)\hss}\vtop{\hbox{}\vspace{-1em}
      \includegraphics[width=0.7\linewidth]{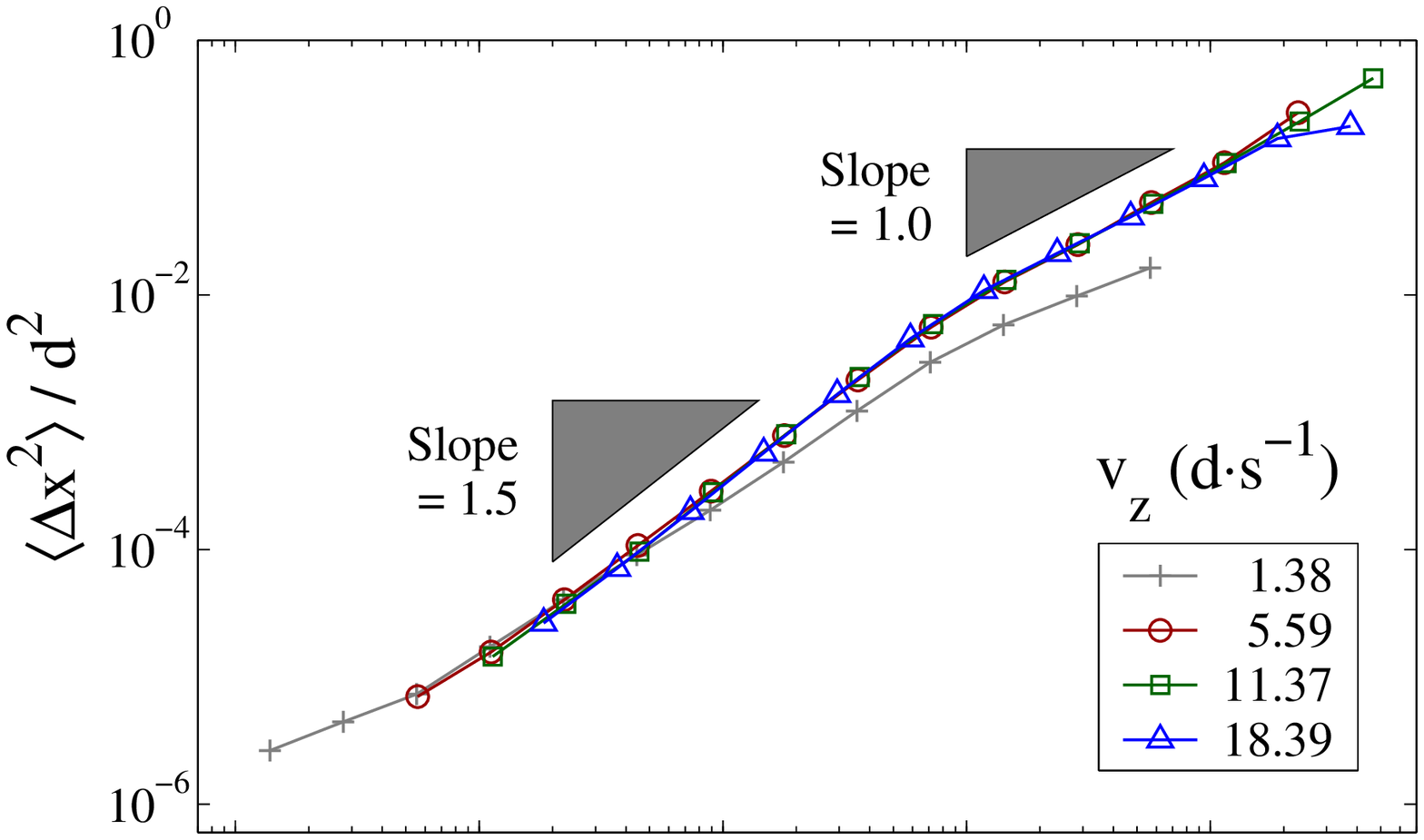}}}
  \hbox{\hbox to 0pt{\qquad(b)\hss}\vtop{\hbox{}\vspace{-1em}
      \includegraphics[width=0.7\linewidth]{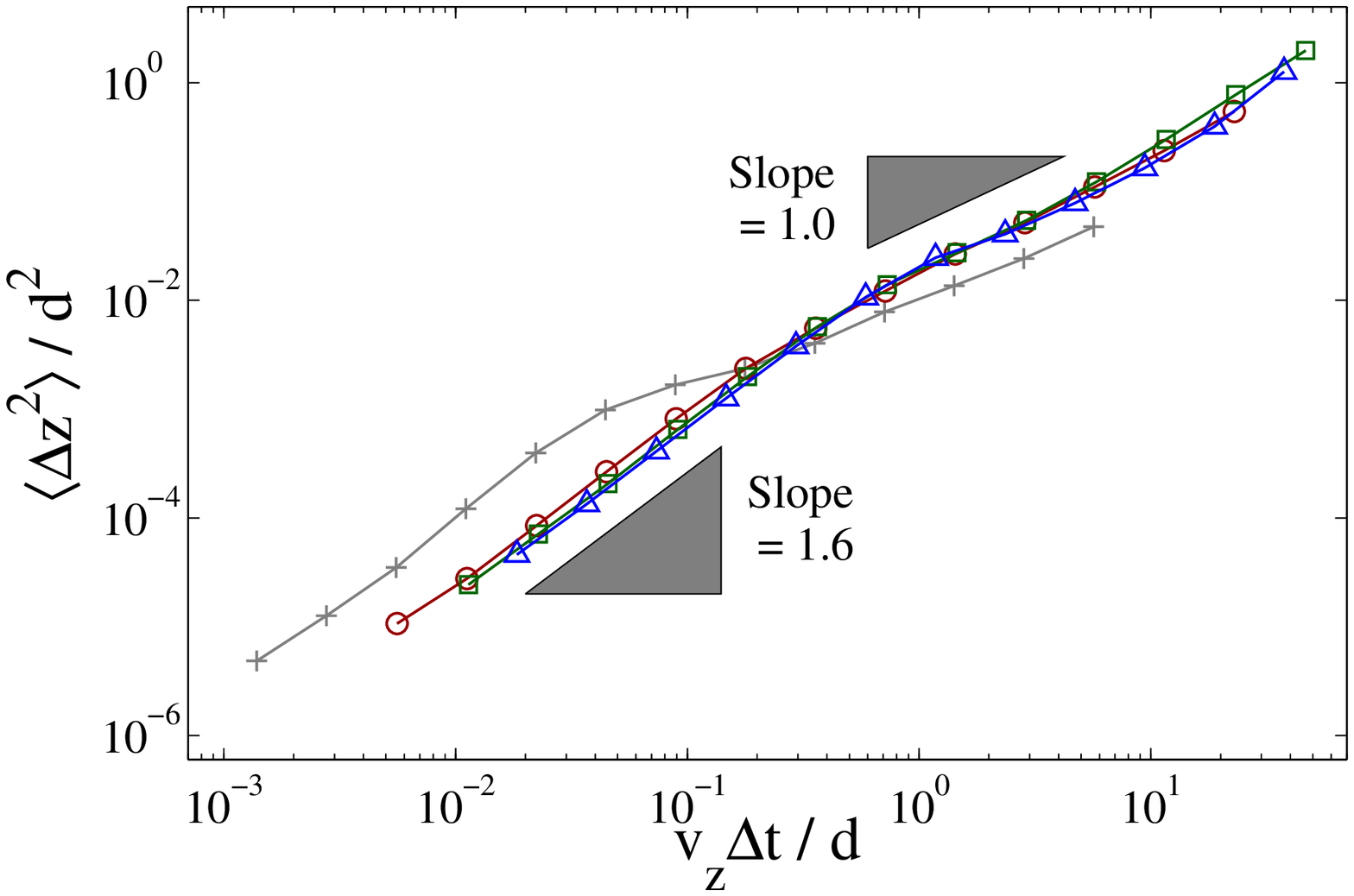}}}
  \hbox{\hbox to 0pt{\qquad(c)\hss}\vtop{\hbox{}\vspace{-1em}
      \includegraphics[width=0.7\linewidth]{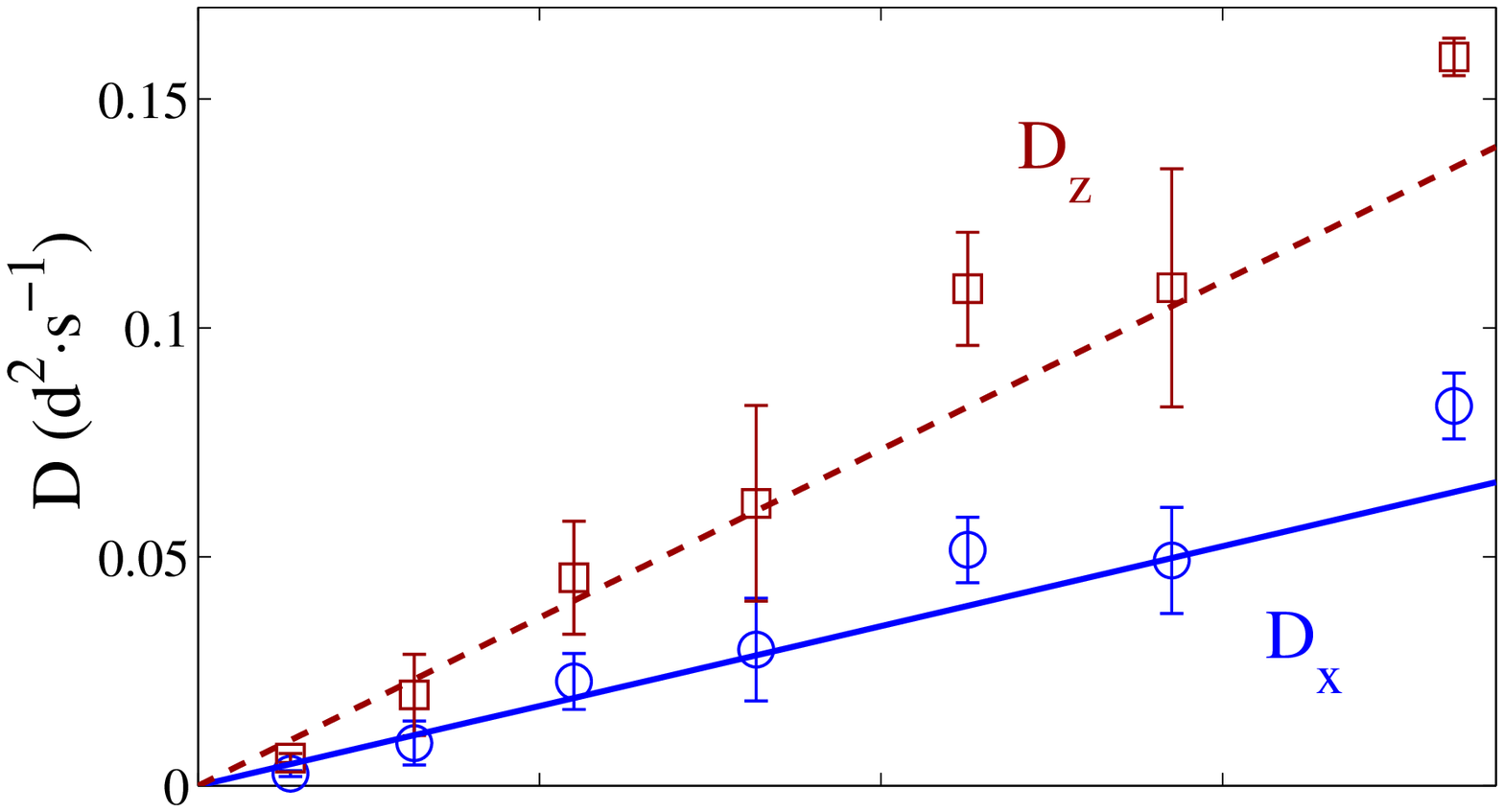}}}
  \hbox{\hbox to 0pt{\qquad(d)\hss}\vtop{\hbox{}\vspace{-1em}
      \includegraphics[width=0.7\linewidth]{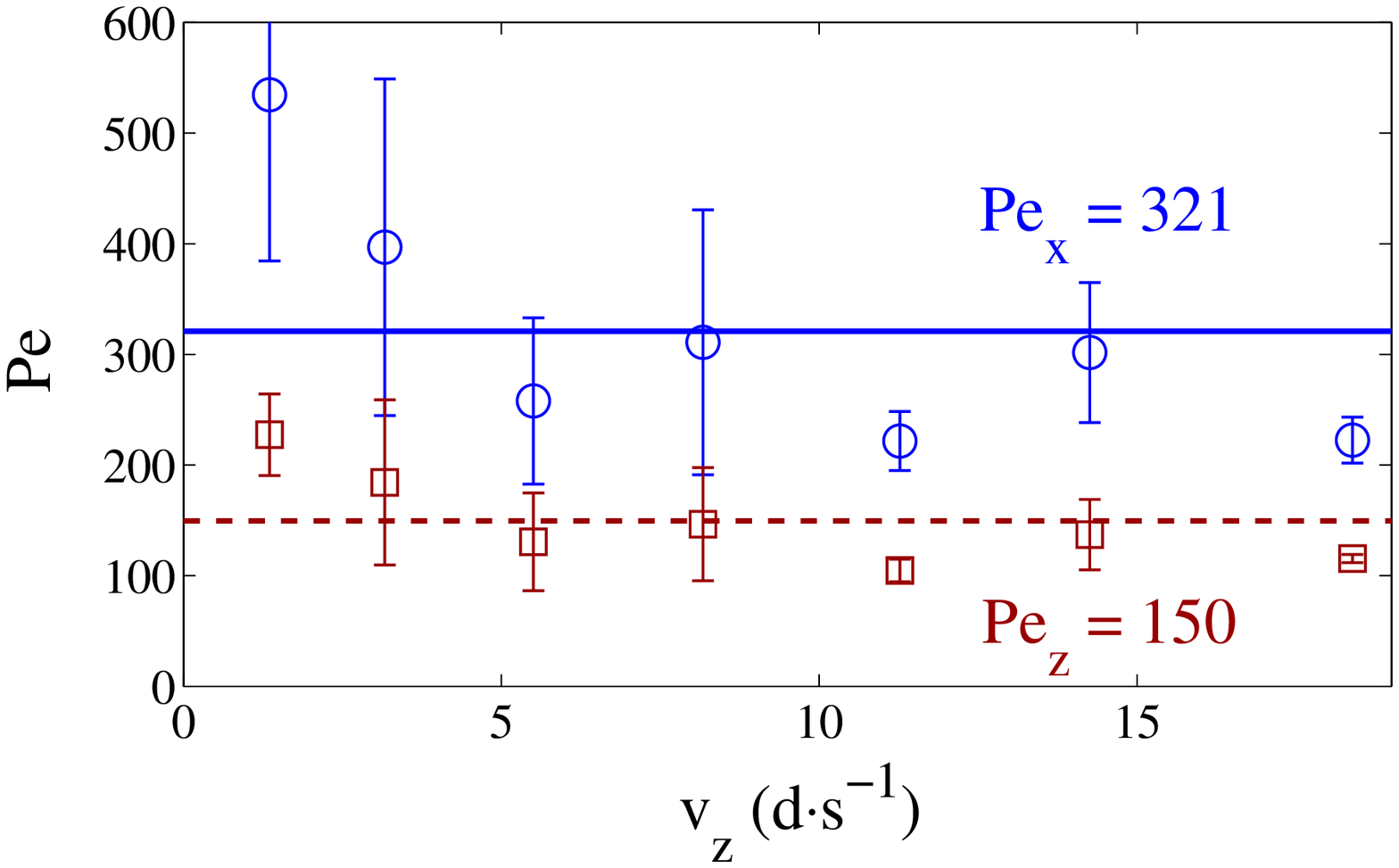}}}
  \caption{\label{Fig3}
    (a)--(b) Mean squared horizontal ($\dxsq$) and vertical ($\dzsq$)
    displacements versus mean distance dropped, which collapse onto a single
    curve for different flow speeds, $\vz$, except for the smallest where 
    the flow is intermittent. (c)--(d) Diffusion coefficients ($D$) and 
    P\'eclet numbers ($\Pe$) in the horizontal ($x$) and vertical ($z$) 
    directions versus $\vz\/$.}
\end{figure}

Looking again at Fig.~\ref{Fig2}(a), it seems that the large
fluctuations in particle displacements would be reduced by
coarse-graining in time, perhaps enough to recover standard Gaussian
statistics. Indeed, as shown in Fig.~\ref{Fig2}(d), the normalized
kurtosis, $\kappa_x = \dxqd/3\dxsq^2\/ - 1$, which measures how the
shape of the distribution of $\dx$ differs from a Gaussian, 
decreases toward zero as $\vz\dt$ increases. (The data fluctuates 
somewhat for $\vz = 1.38 d/$sec, presumably due to intermittency.)  
This suggests a transition from super to normal diffusion.

As shown in Fig.~\ref{Fig3}(a)--(b), the scaling of the mean-square
displacements does, in fact, change from super-diffusive, $\dxsq
\propto \dt^{1.5}\/$ and $\dzsq \propto \dt^{1.6}\/$, to diffusive,
$\dxsq \propto \dzsq \propto \dt$. 
The normal diffusion in long time scales is consistent 
with previous studies of dense drainage where particles were
tracked with lower time resolution~\cite{hsiau93, natarajan95}.
Curiously, the super-diffusion is slower than in the usual case 
of ballistic transport in fluids, $\dxsq\propto \dzsq \propto \dt^2$, 
where particles move freely between collisions. Sub-ballistic scaling 
and non-Gaussian statistics at short times suggest that dense granular 
flows differ from classical fluid flows.

The differences become even more obvious upon varying the flow
rate. In a fluid, this causes a linear increase in $\Pe$ because the
mean flow has no affect on molecular diffusion due to thermal
fluctuations. Here, as shown in Fig.~\ref{Fig3}(c)--(d), the measured
diffusion coefficients, e.g. $D_x = \lim_{\dt\rightarrow\infty}
\dxsq/2\dt$, are actually proportional to the flow speed (with $D_z \approx
2.1 D_x$, consistent with the discussion above), so the P\'eclet
numbers, $\Pe_x = \vz d/D_x \approx 321$ and $\Pe_z = \vz d/D_x
\approx 150$, are roughly constant. This suggests that diffusion and
advection are caused by the same physical mechanism, such as a passing
void. The measured P\'eclet numbers, however, are two orders of
magnitude larger than predicted by the void model.

Since $D_x, D_z \propto \vz$, we plot the mean square displacements
versus the mean distance dropped in the laboratory frame, $\vz\dt$.
Remarkably, as shown in Fig.~\ref{Fig3}(a)--(b), this collapses
all of our data for different flow speeds onto a single curve, not
only in the diffusive regime, but also in the super-diffusive
regime. (The data for the smallest flow speed again differs somewhat.)
A smooth crossover from super to normal diffusion occurs
after particles have fallen roughly one particle diameter.

\begin{figure}
  \hbox{\hbox to 0pt{\qquad(a)\hss}\vtop{\hbox{}\vspace{-1em}
      \includegraphics[width=0.7\linewidth]{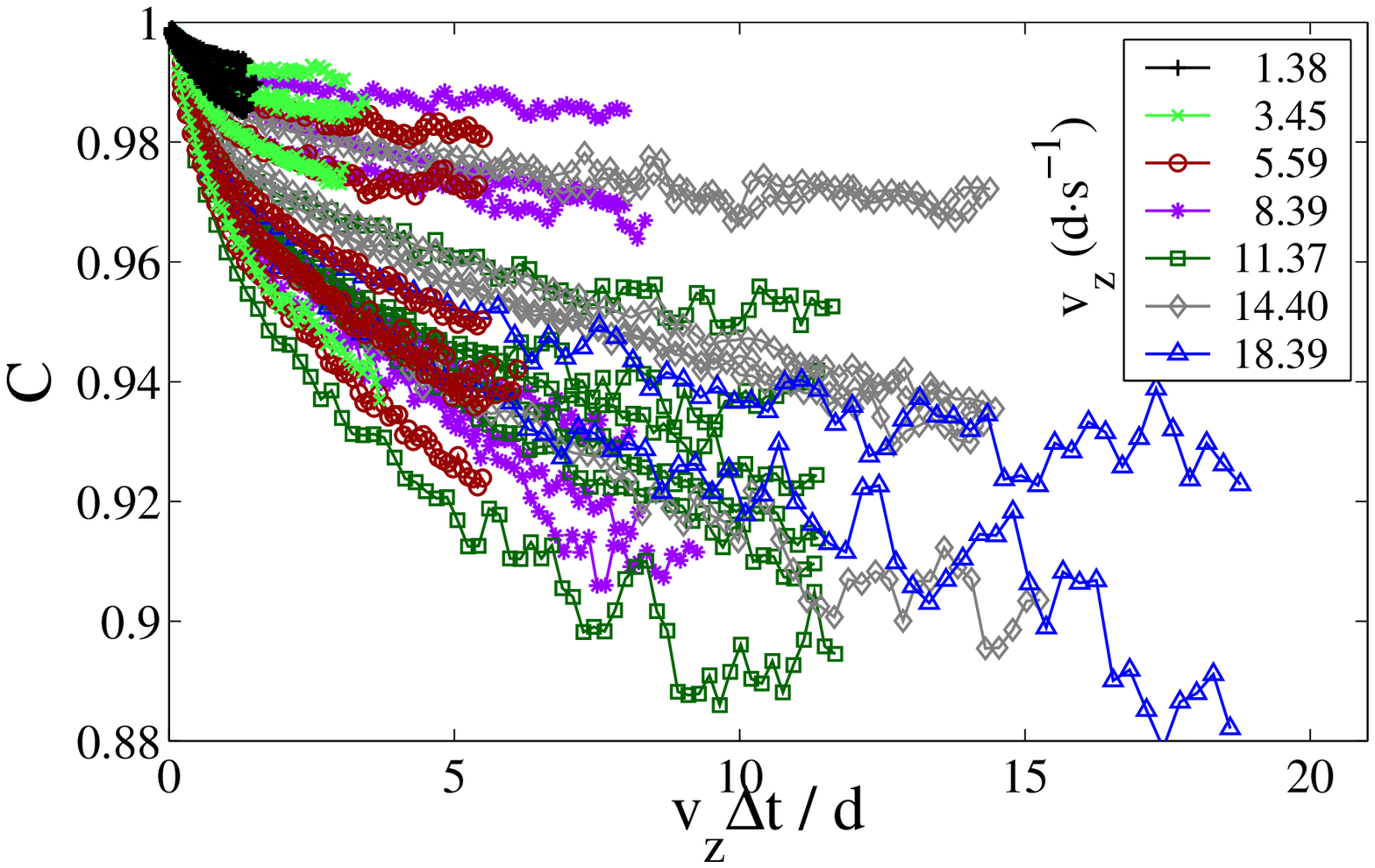}}}
  \hbox{\hbox to 0pt{\qquad(b)\hss}\vtop{\hbox{}\vspace{-1em}
      \includegraphics[width=0.7\linewidth]{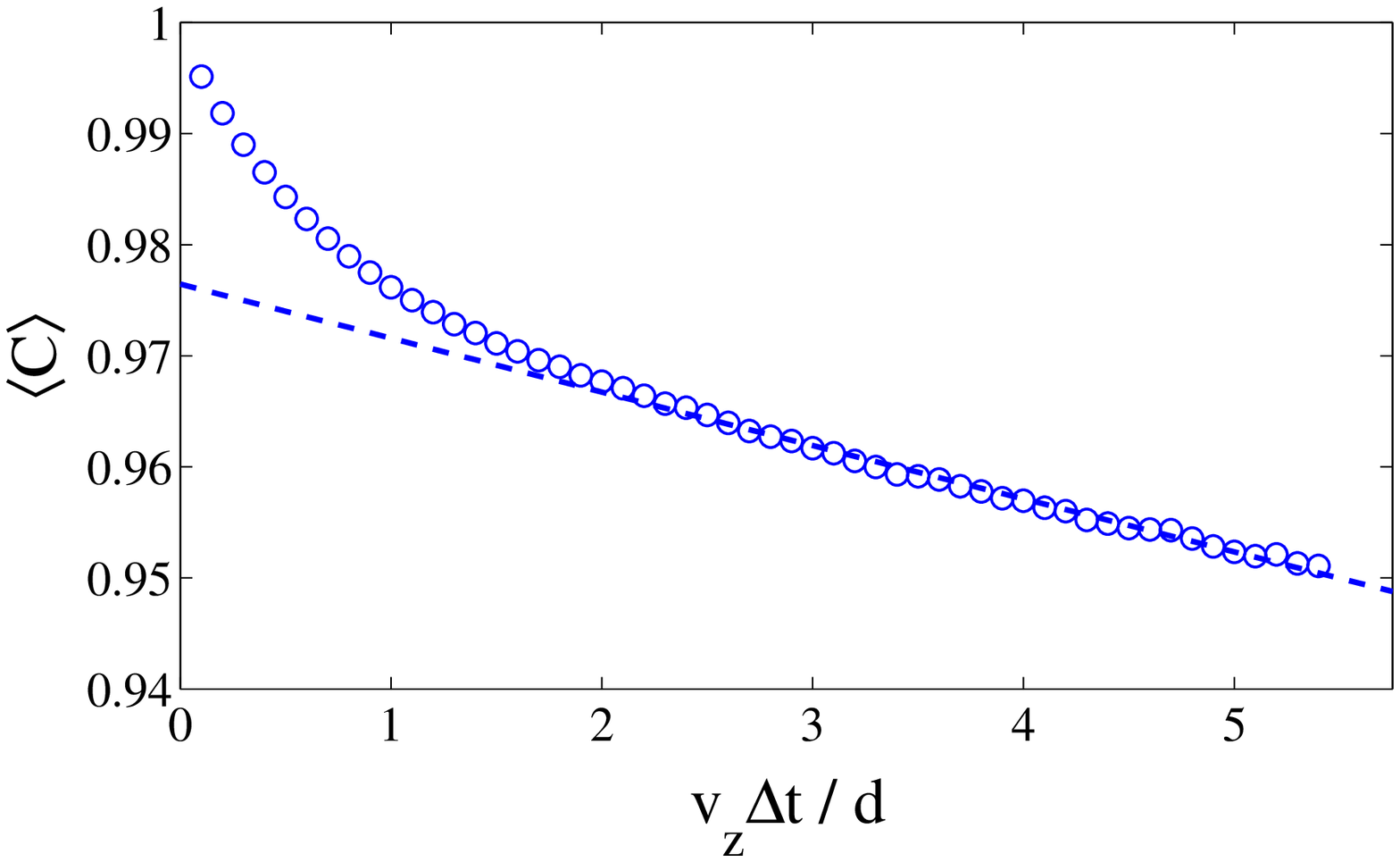}}}
  \caption{\label{Fig5} (a) The topological cage correlation function,
    $C(\dt)\/$, versus the mean distance fallen, $\vz\dt$, and (b) the
    average over all experiments in the continuous-flow regime ($W\ge
    16\;\mathrm{mm}$), compared with an
    exponential fit in the diffusive regime (dotted line).}
\end{figure}

Although advection dominates particle dynamics ($\Pe \gg 1$),
diffusion causes a gradual rearrangement of the `cage' of nearest
neighbors. To investigate this mixing directly, we measure the
topological correlation function, $C(\dt)$, defined as the fraction of
nearest neighbor pairs preserved from times $t$ to $t + \dt$, averaged
over all $t$. We chose the cutoff for a nearest neighbor, $1.5d$, as
the first minimum of the radial distribution function.  (which yields
coordinations near 0.59).  As shown in Fig.~\ref{Fig5}(a), 
the data for $C(\dt)$ collapses when plotted versus $\vz \dt$, in the
sense that no systematic dependence on $\vz$ is observed (except
perhaps for the smallest orifice widths), so in Fig.~\ref{Fig5}(b) we
plot the average over all experiments in the continuous flow regime
($W\ge 16\;\mathrm{mm}$ or $\vz\ge 5.59\;d/sec$).

The cage correlation function in Fig.~\ref{Fig5}(b) exhibits a clear
crossover, which closely parallels the ones for mean displacements in
Fig.~\ref{Fig3}(a)--(b). In the superdiffusive regime, $C(\dt)$
decreases fairly quickly (with a decay length of roughly $20d$), but
after falling more than one particle diameter the rate of decrease
(neighbor loss) slows considerably.  Since the topology remains more
than $90\%$ intact within the observation window, the precise form of
the long-distance decay is uncertain, but a least-squares exponential
fit, $\langle C\rangle \sim 0.976\exp(-\vz\dt/200d)$, yields a ``cage
breaking length'' of $200d$.  This gives direct evidence that the flow
is characterized by long-lasting contacts, as is obvious to the naked
eye.  It also firmly rejects the void model because any void-particle
exchange removes roughly half of the neighbors of a particle as it
falls by only one diameter.

To counter the argument that a collisional regime may exist below the
experimental resolution ($\dt \ll 1$ ms), we show that this is
inconsistent with the fact that diffusion and mixing depend only on
geometry (Figs.~\ref{Fig3}--\ref{Fig5}). In the standard model of a
collisional gas, a particle dropping a distance, $L$, experiences an
average of $N$ collisions which must dissipate its gravitational
potential energy: $m\,g\,L = (1/2)\,N\,(1-e^2)\,m\,\vrel^2\/$, where
$m\/$ is the mass, $g$ the gravitational acceleration, $e\/$ the
restitution coefficient, and $\vrel$ the mean relative velocity. To be
consistent with our data, $N$ should depend on $L$, but not the flow
speed, $\vz$. Although $\vrel$ is unknown, we can make two estimates
--- both of which lead to a contradiction.  The first starts from the
natural formula, $\vrel^2 \approx (\dxsq+\dzsq)/\dt^2$ with fixed $\dt
= 1$ ms, which suggests $\vrel \propto \vz^{0.8}\/$ by looking at the
initial slope of $\dzsq$ in Fig.~\ref{Fig3}(b). The second follows
from direct measurements~\cite{menon97} of $\vrel\/$ yielding $\vrel
\propto\vz^{2/3}$. In either case, $N$ would not be constant. (Note
that $(1-e^2)\/$ would typically correlate with $\vrel^2\/$, so
velocity-dependent restitution cannot compensate for the changes in
$\vrel\/$.) 

More generally, in slow granular flows it seems that 
``granular temperature'' may not be a
useful concept. Figures \ref{Fig3} and \ref{Fig5} clearly show
that fluctuations depend only on the distance fallen, and yet any
notion of temperature should increase with the flow speed. The fact that the
nearest-neighbor topology persists for distances comparable to the
system size also seems to cast doubt on the assumption of ergodicity.

Instead, our experimental data suggests that cage rearrangements are
caused by the relaxation of contact networks, as are believed to occur
in Couette cells~\cite{howell99}. Such networks could absorb potential
energy via rolling and sliding neighbors. The breaking of a contact
could cause non-Gaussian fluctuations and small-scale superdiffusion
among the particles in a network, while the gradual destruction of a
network (and reformation of a new one) as a particle falls farther than
its own diameter could cause the observed transition to normal
diffusion. All of these effects are dominated by the geometry of
random close packings, which is not fully understood, even without any
dynamics~\cite{torquato00}.

In summary, we have experimentally investigated particle dynamics in
dense granular flows --- as they occur in silo drainage. Consistent
with the void model, we observe diffusion after drainage by more than
a particle diameter and P\'eclet numbers which are independent of the
flow rate, suggesting that advection and diffusion have the same
physical source. The P\'eclet numbers and cage-breaking lengths,
however, are much larger than predicted, so the question of an
appropriate microscopic model is left open.

\begin{acknowledgments}
We thank A. Samadani and D. Blair for helpful discussions. This work
was supported by the Department of Energy (grant \#DE-FG02-02ER25530),
the Norbert Weiner Research Fund, and NEC Fund at MIT and by the
National Science Foundation (grant \#DMR-9983659) at Clark.
\end{acknowledgments}

\end{document}